\documentclass[a4paper,10pt]{article}
\usepackage[utf8]{inputenc}

\usepackage{amsmath}
\usepackage{subcaption}
\usepackage{multirow}

\DeclareMathAlphabet{\mathpzc}{OT1}{pzc}{m}{it}
\usepackage{graphicx}
\newcommand{\eq}[1]{(\ref{#1})}

\usepackage[top=2.5cm, bottom=2.5cm, left=2.1cm, right=2.1cm]{geometry}

\title{Rebuttal of ``The multicaloric effect in multiferroic materials''}
\author{Ivan A.~Starkov$^1$ and Alexander S.~Starkov$^{2*}$\vspace{2mm}\\
$^1$Nanotechnology Center, St.~Petersburg Academic University,\\ Russian Academy of Sciences, St. Petersburg, 194021, Russia\vspace{2mm}\\ 
$^{2}$National Research University of Information Technologies,\\ Mechanics and Optics, St.~Petersburg, 197101, Russia.\vspace{1mm}\\
	      $^{*}$corresponding author; e-mail: ferroelectrics@ya.ru
}
\begin{document}
\date{}
\maketitle

\noindent

\begin{abstract}
The paper refutes the model and claims published in the Solid State Communications \cite{Vopson_12} as well as elsewhere.
The theoretical approach proposed in ``The multicaloric effect in multiferroic materials'' by Melvin M.~Vopson has a number of inaccuracies and mistakes. 
The Author of \cite{Vopson_12} does not pay necessary attention to the range of applicability of the derived equations and confuses the dependent and independent variables. 
The resulting equations for electrically and magnetically induced multicaloric effects are incorrect and cannot be used for measurement interpretation. 
\end{abstract}

\vspace*{0.5cm}

In August 2012, Solid State Communications published an article by Melvin M.~Vopson entitled ``The multicaloric effect 
in multiferroic materials'' \cite{Vopson_12}. The Author pretends on the theoretical introduction of the multicaloric effect in multiferroics.
However, we feel that the article has several flaws and that the conclusions are unfounded. Melvin M.~Vopson makes contradictory 
assumptions -- in the beginning, the electric and magnetic fields are constant, but afterwards the Author considers them as variable quantities. 
We have provided the following rebuttal to clarify these theoretical issues. For convenience, we keep the notations introduced 
in \cite{Vopson_12}.

In order to quantify the caloric effects, Melvin M.~Vopson uses the differential changes of the entropy in terms of the
differential changes of the applied fields

\begin{equation}
  dS=\cfrac{C}{T}\cdot dT+\left(\cfrac{\partial M}{\partial T}\right)_{H,E}\cdot dH+\left(\cfrac{\partial P}{\partial T}\right)_{E,H}\cdot dE.
\label{eq:1}
\end{equation}
In the case of an adiabatic process $dQ=T \cdot dS=0$, the electrically and magnetically induced multicaloric effects in 
multiferroics are expressed as \cite{Vopson_12}

\begin{subequations}
\begin{align}
 \Delta T_E=-T\cdot\int^{E_f}_{E_i}\cfrac{1}{C_{E,H}}\left[\cfrac{\alpha_e}{\mu_0\chi^m}\cdot\left(\cfrac{\partial M}{\partial T}\right)_{H,E}+\left(\cfrac{\partial P}{\partial T}\right)_{H,E}\right]\cdot dE,\\ 
 \Delta T_H=-T\cdot\int^{H_f}_{H_i}\cfrac{1}{C_{E,H}}\left[\left(\cfrac{\partial M}{\partial T}\right)_{H,E}+\cfrac{\alpha_m}{\mu_0\chi^e}\cdot\left(\cfrac{\partial P}{\partial T}\right)_{H,E}\right]\cdot dH, 
\end{align}
\label{eq:2}
\end{subequations}
For the transition from \eq{eq:1} to \eq{eq:2}, the Author first assumes the linear relationship between $M$, $P$ and their corresponding fields $H$, $E$

\begin{subequations}
\begin{align}
  d M=\mu_0 \chi^m dH\\
  d P=\varepsilon_0 \chi^e dE.
\end{align}
\label{eq:3}
\end{subequations}
Then, considering the linear magneto-electric effect, the relations
\begin{subequations}
\begin{align}
 d M=\alpha_e  dE,\\
 d P= \alpha_m dH. 
\end{align}
\label{eq:4}
\end{subequations}
are used to distinguish between the electrically and magnetically induced coupling.
By formal manipulation of \eq{eq:3} and \eq{eq:4}, Melvin M.~Vopson deduces the equations

\begin{subequations}
\begin{align}
 d H=\alpha_e(\mu_0 \chi^m)^{-1} dE,\\
 d E=\alpha_m(\varepsilon_0 \chi^e)^{-1} dH, 
\end{align}
\label{eq:5}
\end{subequations}
which after the substitution to \eq{eq:1} lead to \eq{eq:2}.

Let us find out under what conditions the resulting equations \eq{eq:2} are correct. Recall that 
the general relations for the linear magnetoelectric effect in the scalar case are \cite{7,17}

\begin{subequations}
\begin{align}
  d M=\mu_0 \chi^m dH+ \alpha_e  dE, \\
  d P=\alpha_m dH+\varepsilon_0 \chi^e dE .  
\end{align}
\label{eq:6}
\end{subequations}
Thus, it is possible to highlight the assumptions necessary for the use of (\ref{eq:3}a,b) and (\ref{eq:4}a,b)
\begin{center}
\begin{tabular}{llcrr}

(\ref{eq:3}a) & $E=$ {const},&	 \multirow{2}{*}{and}	& (\ref{eq:4}a) & $H=$ {const},\\
(\ref{eq:3}b) & $H=$ {const},&				& (\ref{eq:4}b) & $E=$ {const}.\\
\end{tabular}
\end{center}
Since the equations \eq{eq:5} are obtained by combining (\ref{eq:3}a,b) and (\ref{eq:4}a,b), they are correct only for
$H = \mathrm{const}$ and $E = \mathrm{const}$. In this case, both parts of \eq{eq:5} are identically equal to 0. Therefore, 
all subsequent formulas remain valid for the simultaneously constant electric and magnetic fields. Moreover, when applicable,
they turn into the trivial equality $0 = 0$. In other words, they do not carry any information.
Besides the fact that Melvin M.~Vopson does not pay attention to the conditions of applicability of the derived equations, 
he makes another blunder. In \eq{eq:3} and \eq{eq:4} the variables $E$ and $H$ are independent, but in \eq{eq:5} 
they are linked by linear relations, i.e.~already dependent variables.

From the experimental point of view, the equations \eq{eq:2} mean an increase in magnitude of the caloric effect with decreasing dielectric or magnetic susceptibility.
At the same time, as it was well established \cite{1,8}, for the main caloric effects the magnitude of the effect increases with increasing dielectric or magnetic susceptibility.
In addition, the expressions \eq{eq:2} are incorrect as obtained by taking out the variable temperature $T$ of the integral.
The Author of \cite{Vopson_12} tries to describe the temperature change of the system produced by an external electric and/or magnetic field.
That is, from the problem statement, temperature is a function of $E$ and $H$ and cannot be pull out in front of the integrals as in \eq{eq:2}. 
It is also worth to mention that the model \cite{Vopson_12} considers the magnetocaloric coefficient to be independent on temperature. 
This assumption does not work for artificial composite multiferroics, which results in the inapplicability of the whole approach 
for such a class of materials.

On the basis of the reported errors, it is possible to argue that the conclusion ``the theoretical work developed in this paper allows the unique derivation of the
magneto-electric coupling coefficient of multiferroics from thermal measurements'' \cite{Vopson_12} is far from reality.
In  summary, we hope that our critical remarks will help to contribute to the further development of the multicaloric effect
and will be taken into consideration in the subsequent investigations.

%

\end{document}